\documentclass[preprint,pra,amssymb,amsmath,showpacs,floatfix]{revtex4}

\usepackage{graphicx}
\usepackage{amssymb}
\usepackage{amsmath}

\newcommand{\be}{\begin{equation}}
\newcommand{\ee}{\end{equation}}
\newcommand{\bea}{\begin{eqnarray}}
\newcommand{\eea}{\end{eqnarray}}

\begin{document}

\title{ Alternative Perspective on Quantum Tunneling and Instantons }

\author{F.~Paradis$^{a}$, H.~Kr\"{o}ger$^{a}$$\footnote{Corresponding author, Email: hkroger@phy.ulaval.ca}$, G.~Melkonyan$^{a}$, K.J.M.~Moriarty$^{b}$  }

\affiliation{
$^{a}$ {\small\sl D\'{e}partement de Physique, Universit\'{e} Laval, Qu\'{e}bec, Qu\'{e}bec G1K 7P4, Canada} \\
$^{b}$ {\small\sl Department of Mathematics, Statistics
and Computer Science, Dalhousie University, Halifax N.S. B3H 3J5, Canada}
}

\begin{abstract}
We present a new way to compute and interpret quantum tunneling in
a 1-D double-well potential. For large transition time we show
that the quantum action functional gives an analytical expression
for tunneling amplitudes. This has been confirmed by numerical
simulations giving relative errors in the order of
$10^{-5}$. In contrast to the classical potential, the
quantum potential has a triple-well if the classical
wells are deep enough. Its minima are located at the position of
extrema of the ground state wave function. The striking feature is
that a single trajectory with a double
instanton reproduces the tunneling amplitude. This is in
contrast to the standard instanton approach, where infinitely many
instantons and anti-instatons have to be taken into account. The
quantum action functional is valid in the deep quantum regime
in contrast to the semi-classical regime where the standard instanton
approach holds. We compare both approaches via numerical
simulations. While the standard instanton picture describes only
the transition between potential minima of equal depth, the
quantum action may give rise to instantons also for asymmetric potential
minima. Such case is illustrated by an example.
\end{abstract}


\maketitle



\section{Introduction}
\label{sec:Intro} Tunneling is a characteristic feature of quantum
physics, having no counter part in classical physics. Instantons
are known to be intimately connected to tunneling. The physics of
instantons and its relation to tunneling have been discussed in
Refs.\cite{Coleman,Rajaraman} and the role of instantons in QCD
has been reviewed in Ref.\cite{Schaefer}. Tunneling effects and
the use of instantons appear in many different areas of physics,
such as inflationary scenarios and formation of galaxies
\cite{Staro79,Khlo98,Kolb91,Q2}, hot and dense nuclear matter
\cite{Shur88}, neutrino oscillations
\cite{NeutrOsc1,NeutrOsc2,NeutrOsc3}, condensed matter physics
(SQUIDs) \cite{Friedman00,Averin00}, quantum computers based on
superconductors \cite{Gulian1,Gulian2}, dynamical tunneling
\cite{Raizen, Hensinger} and chemistry (chemical bindings).

The standard instanton picture is valid in the semi-classical
regime. Infinitely many instantons and anti-instantons contribute
to give the tunneling amplitude. In this work we consider the
opposite regime, i.e. the deep quantum regime. We use the
concept of the quantum action, being a kind of effective action,
which takes into account quantum effects via tuned action
parameters. The action is computed from the ground state of the
system. In the limit of large imaginary time, the ground state wave function determines the shape of the quantum potential, which together with a corresponding quantum mass determines the quantum action.  
The quantum action functional then gives the exact
tunneling amplitudes. The shape of the resulting quantum potential
is different from the classical double well potential, i.e. it
exhibits a triple well structure. We find that a double instanton
(resp. anti-instanton) is necessary and sufficient to reproduce
exactly the tunneling amplitude. The standard instanton approach holds for oscillator-like potentials with deep wells and high barriers. In contrast to that, the quantum action approach holds when the potential is shallow and the physics is dominated by the ground state properties (Feynman-Kac limit), i.e. in the deep quantum regime. In this sense the quantum action
functional is a method complementary to the semi-classical instanton approach.

Another approach in constructing an effective classical potential has been proposed by Feynman and Kleinert \cite{Feynman86}. Though similar to the quantum action in its physical goal and motivation, it differs by its definition. When applied to a classical double well potential \cite{Janke87} it gives an effective potential different from the quantum potential. In particular, the quantum action  generates a triple-well potential with degenerate minima (all of equal depth), which is not the case for the effective classical potential.  

In Sect.\ref{sec:ModelAndAction} we present the tunneling model and
the construction of the quantum action functional.
Sect.\ref{sec:NumSim} presents numerical results on how the
quantum action functional has been calculated and how well it fits
the transition amplitude. In Sect.\ref{sec:CompInst} we compare
numerical results from the standard instanton approach with those
from the quantum action functional. We briefly discuss the use of
the quantum action method for asymmetric double-well potentials in
Sec.\ref{sec:tilted}. Finally, Sect.\ref{sec:Discuss} gives a
discussion and Sect.\ref{sec:Summary} a summary.

\section{Model and its quantum action}
\label{sec:ModelAndAction}
\subsection{Quantum mechanical tunneling model}
\label{sec:Model}
\begin{figure*}[ht]
\begin{center}
\includegraphics[scale=0.70,angle=0]{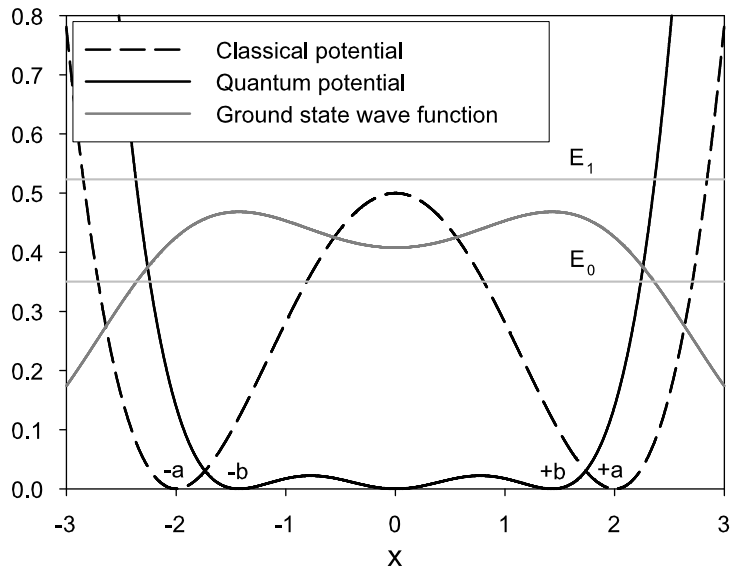}
\end{center}
\caption{Classical double well potential for
$\lambda=\frac{1}{32}$ (minima at $-a$, $+a$), the ground state
wave function and two lowest energy levels. The quantum potential
$\tilde{V}-\tilde{V}_0$ (multiplied by a scale factor 1.7331)
displays triple wells (minima at $-b$, $0$, $+b$). All quantities
are in dimensionless units.}
\label{fig:PotL1-32}
\end{figure*}

\noindent Let us consider in 1-D a classical Hamiltonian system,
\begin{equation}
\label{eq:Hamilton}
H = \frac{p^{2}}{2m} + V(x) ~ ,
\end{equation}
with a potential of double-well shape given by
\begin{equation}
\label{eq:Potential}
V(x) \equiv \lambda \Big( x^2 - \frac{1}{8\lambda} \Big)^{2} =
\lambda x^4-\frac{x^2}{4}+\frac{1}{64\lambda} ~ .
\end{equation}
For simplicity, we use throughout $\hbar = c = m = 1$, which makes all physicals units dimensionless. The potential minima are located at
$\pm a$, $a=\frac{1}{\sqrt{8\lambda}}$.
A potential barrier of height $B =
\frac{1}{64\lambda}$ is located at $x=0$ (see Fig.[\ref{fig:PotL1-32}]).
The potential parameters were chosen such that the natural frequency
of the oscillations at the bottom of each well is $\omega = 1$ for
all $\lambda$. The potential is of confinement-type, i.e. tends to
infinity for large $|x|$. Hence the quantum system has a discrete
bound state spectrum (no scattering states) of energies $E_{n}$,
$n=0,1,2,\cdots$. Depending on the height of the potential barrier
$B$, the following physical situations may occur: (i) $B < E_{0}$.
The barrier is lower than the ground state energy. Then the ground
state wave function has a single hump at the center. (ii) $E_{0} <
B$. The ground state energy is lower than the barrier and the
ground state wave function displays a double hump. This case is
sketched in Fig.[\ref{fig:PotL1-32}]. (iii) $E_{0} < E_{1} <
\cdots E_{k} < B$. The barrier is higher than the first $k+1$
bound state energies. Quantum tunneling in the proper sense occurs
in cases (ii) and (iii). The transition between regions (i) and
(ii) occurs at $\lambda = 5.34581336\cdot 10^{-2}$. In all plots
involving $\lambda$, a vertical gray line represents this
boundary. The parameter $\lambda$ controls the barrier height and
the property of the system to be located in the semi-classical
regime. Large $\lambda$ means a low barrier (quantum regime) while
small $\lambda$ represents two deep (almost decoupled) wells
(semi-classical regime). The tunneling amplitude is given by the
(imaginary time) transition matrix element corresponding to the
transition from one minimum to the other,
\begin{equation}
\langle a | \exp[- HT/\hbar] | -a \rangle ~ .
\end{equation}

\subsection{Construction of the quantum action for large transition time}
\label{sec:ConstQAction}
The quantum action has been introduced in Ref.\cite{Q1}.
It is a local action, like the classical action, given by
\begin{equation}
\label{eq:DefQAction}
\tilde{S}[x] = \int_{0}^{T} dt ~ \frac{\tilde{m}}{2} \dot{x}^{2} - \tilde{V}(x) ~ ,
\end{equation}
where $\tilde{m}$ is called the quantum mass and $\tilde{V}(x)$
the quantum potential.
The quantum action functional is defined as a parametrisation of the quantum transition amplitude $G(x_{fi},t=T;x_{in},t=0)$ for arbitrary transition time $T$ in the following way
\begin{eqnarray}
\label{eq:DefQFunctional}
&& G(x_{fi},t=T;x_{in},t=0) = \sum_{trajectories} \tilde{Z} \exp[ i \tilde{S}[\tilde{x}_{traj}] /\hbar ] ~ ,
\nonumber \\
&& \tilde{x}_{traj} : \delta \tilde{S}[\tilde{x}_{traj}] = 0 ~ .
\end{eqnarray}
Here $\tilde{x}_{traj}$ denotes a trajectory (stationary point) of the action $\tilde{S}$ going from boundary point $(x_{in},t=0)$ to $(x_{fi},t=T)$.
There may be several such trajectories.

Tunneling involves imaginary time $(t \to -i t)$. Moreover, the instanton picture of tunneling is usually considered in the limit of large
transition time $(T \to \infty)$. Hence let us consider from now on time to be imaginary and transition time $T$ to become large (Feynman-Kac limit).
In this limit the quantum action functional has been proven to exist,
and to give an exact parametrisation of transition amplitudes by taking into account only a single trajectory \cite{Q4}
\begin{equation}
\label{eq:DefQFunctionalEucl}
G_{Eucl}(x_{fi},t=T;x_{in},t=0) \longrightarrow_{T \to \infty} \tilde{Z}_{Eucl} \exp[ - \tilde{S}_{Eucl}[\tilde{x}_{Eucl}^{traj}] /\hbar ] ~ ,
\end{equation}
where
\begin{equation}
\label{eq:DefQActionEucl}
\tilde{S}_{Eucl}[x] = \int_{0}^{T} dt ~ \frac{\tilde{m}}{2} \dot{x}^{2} + \tilde{V}(x) ~ ,
\end{equation}
denotes the Euclidean action (following physics conventions, we dropped the overall minus sign occuring in the action at imaginary time. It will reappear in $\exp[-\tilde{S}_{Eucl}]$).
Here $\tilde{x}_{Eucl}^{traj}$ is the trajectory which makes $\tilde{S}_{Eucl}$ stationary.
We use the notation $\tilde{\Sigma} = \tilde{S}[\tilde{x}_{Eucl}^{traj}]$.
It should be pointed out that, although being a stationary point of the action $\tilde{S}_{Eucl}$, the trajectory $\tilde{x}_{Eucl}^{traj}$ does not necessarily always minimize the quantum action. Below we will show which particular trajectory must be taken in order to represent the propagator. From now on we drop the subscript Euclidean.

Also in this limit, analytical relations exist between the
classical potential, the quantum potential and wave functions
\cite{Q5}. For example the following relation was established
\cite{Q4},
\begin{equation}
\label{eq:DefQPot}
\frac{2 \tilde{m}}{\hbar^{2}} (\tilde{V}(x) -
\tilde{V}_{0}) = \left( \frac{d \psi_{gr}(x)/dx}{\psi_{gr}(x) }
\right)^{2} = \left( \frac{d \ln \psi_{gr}(x)}{dx} \right)^{2} ~ .
\end{equation}
Here $\tilde{V}_{0}$ denotes the minimum value of the quantum
potential.
Those results have been established on the assumption of a non-degenerate ground state and a confining potential with a single minimum. Tunneling involves potentials with multiple (possibly degenerate) minima. Therefore, we take here Eq.(\ref{eq:DefQPot}) as starting point and aim to construct an expression which generalizes that of Eq.(\ref{eq:DefQFunctionalEucl}).

From the Schr\"odinger equation one can compute the ground state wave function
$\psi_{gr}(x)$ and via Eq.(\ref{eq:DefQPot}) obtain the quantum potential (times the quantum mass). A plot of the quantum potential
$\tilde{V}(x) - \tilde{V}_{0}$
(up to a multiplicative factor) is shown in Fig.[\ref{fig:PotL1-32}]. The following observations can be made. First, the quantum potential, like the classical potential, is of confining type, i.e. it goes to infinity
for large $|x|$. It also displays the same dependence in
$x^4$ for large $x$. Second, there is a marked difference between
the shape of the classical potential and the quantum potential.
The former has two wells, but the latter can have one, two
or, in the case shown in Fig.[\ref{fig:PotL1-32}], three wells,
located at $-b$, $0$ and $+b$. Moreover, Eq.(\ref{eq:DefQPot}) shows that
$\tilde{V}(x)=\tilde{V}_{0}$, i.e. the quantum potential reaches a
minimum, whenever the ground state wave function has derivative
zero, i.e. whenever it reaches a maximum or a minimum. This
correspondence between the extrema of the ground state wave
function and the quantum action is a property not shared by the
classical potential. Finally, the parity symmetry of classical
potential is maintained by the quantum potential also.

Integration of Eq.(\ref{eq:DefQPot}) allows to express the ground state wave function in terms of the quantum mass and potential,
\begin{equation}
\label{eq:WaveFct}
\psi_{gr}(x) = Z \exp\left[ \pm \int_{x_{0}}^{x} ds \sqrt{\frac{2\tilde{m}}{\hbar^{2}} (\tilde{V}(s) - \tilde{V}_{0}) } \right] ~ ,
\end{equation}
where $Z$ is some integration constant. In this section,
we will consider only the tunneling regime, where the quantum
potential shows a triple well structure --- other cases are simpler.
The question arises:
Which is the physically valid sign in the exponent? The answer can
be found by looking at the shape of $\psi_{gr}(x)$ and
$\tilde{V}(x)$. In the regime $-\infty < x < -b$ (regime I),
$\psi_{gr}(x)$ decreases when $x$ goes from $-b$ to $-\infty$. In
the regime $-b < x < 0$ (regime II), $\psi_{gr}(x)$ increases when
$x$ goes from zero to $-b$. In the regime $0 < x < +b$ (regime
III), $\psi_{gr}(x)$ increases when $x$ goes from zero to $+b$. In
the regime $+b < x < +\infty$ (regime IV), $\psi_{gr}(x)$
decreases when $x$ goes from $+b$ to $+\infty$. This behavior
requires the following choice of signs,
\begin{eqnarray}
\psi_{gr}(x) =
\left\{
\begin{array}{cr}
Z_{I} \exp \left[ - \int_{x}^{-b} ds \sqrt{\frac{2\tilde{m}}{\hbar^{2}} (\tilde{V}(s) - \tilde{V}_{0}) } \right] ~&~ -\infty < x < -b
\\
\\
Z_{II} \exp \left[ + \int_{x}^{0} ds \sqrt{\frac{2\tilde{m}}{\hbar^{2}} (\tilde{V}(s) - \tilde{V}_{0}) } \right] ~&~ -b < x < 0
\\
\\
Z_{III} \exp \left[ + \int_{0}^{x} ds \sqrt{\frac{2\tilde{m}}{\hbar^{2}} (\tilde{V}(s) - \tilde{V}_{0}) } \right] ~&~ 0 < x < +b
\\
\\
Z_{IV} \exp \left[ - \int_{+b}^{x} ds \sqrt{\frac{2\tilde{m}}{\hbar^{2}} (\tilde{V}(s) - \tilde{V}_{0}) } \right] ~&~ +b < x < \infty
\end{array}
\right\} ~ .
\end{eqnarray}
At the boundary of the regimes, the wave function has to be continuous. This implies relations between the factors $Z_{I}, \dots, Z_{IV}$,
\begin{eqnarray}
\label{eq:Zrelations}
\begin{array}{rccl}
\mbox{x = -b:} &
\psi_{gr}^{I}(-b) = \psi_{gr}^{II}(-b)
&\Rightarrow&
Z_{I} = Z_{II} ~ \exp \left[ + \int_{-b}^{0} ds \sqrt{\frac{2\tilde{m}}{\hbar^{2}} (\tilde{V}(s) - \tilde{V}_{0}) } \right]
\nonumber \\
\mbox{x = 0:} &
\psi_{gr}^{II}(0) = \psi_{gr}^{III}(0)
&\Rightarrow&
Z_{II} = Z_{III}
\nonumber \\
\mbox{x = +b:} &
\psi_{gr}^{III}(+b) = \psi_{gr}^{IV}(+b)
&\Rightarrow&
Z_{III} ~ \exp \left[ + \int_{0}^{+b} ds \sqrt{\frac{2\tilde{m}}{\hbar^{2}} (\tilde{V}(s) - \tilde{V}_{0}) } \right]
= Z_{IV} ~ .
\end{array}
\end{eqnarray}
Because the ground state wave function $\psi_{gr}(x)$
and the quantum potential, Eq.(\ref{eq:DefQPot}), are parity symmetric,
the previous equation reduces to
\begin{eqnarray}
\label{eq:ContCond}
&& Z_{II} = Z_{III} \equiv Z
\nonumber \\
&& Z_{I} = Z_{IV} = Z J
\nonumber \\
&& J \equiv \exp\ \left[ + \int_{0}^{+b} ds \sqrt{\frac{2\tilde{m}}{\hbar^{2}} (\tilde{V}(s) - \tilde{V}_{0}) } \right] ~ .
\end{eqnarray}
$Z$ is a factor which normalises the wave function to unity, and
$\log J$ will turn out to be the quantum action of the quantum instanton.

Taking into account those continuity conditions, Eq.(\ref{eq:ContCond}), the wave function $\psi_{gr}(x)$ can be expressed as
\begin{eqnarray}
\label{eq:GroundState}
\psi_{gr}(x) = Z
\left\{
\begin{array}{lr}
\exp \left[
- \int_{x}^{-b} ds \sqrt{\frac{2\tilde{m}}{\hbar^{2}} (\tilde{V}(s) - \tilde{V}_{0}) }
+ \int_{-b}^{0} ds \sqrt{\frac{2\tilde{m}}{\hbar^{2}} (\tilde{V}(s) - \tilde{V}_{0}) }
\right] ~&~ -\infty < x < -b
\\
\\
\exp \left[ + \int_{x}^{0} ds \sqrt{\frac{2\tilde{m}}{\hbar^{2}} (\tilde{V}(s) - \tilde{V}_{0}) } \right] ~&~ -b < x < 0
\\
\\
\exp \left[ + \int_{0}^{x} ds \sqrt{\frac{2\tilde{m}}{\hbar^{2}} (\tilde{V}(s) - \tilde{V}_{0}) } \right] ~&~ 0 < x < +b
\\
\\
\exp\ \left[
+ \int_{0}^{+b} ds \sqrt{\frac{2\tilde{m}}{\hbar^{2}} (\tilde{V}(s) - \tilde{V}_{0}) }
- \int_{+b}^{x} ds \sqrt{\frac{2\tilde{m}}{\hbar^{2}} (\tilde{V}(s) - \tilde{V}_{0}) }
\right] ~&~ +b < x < +\infty
\end{array}
\right\} ~ .
\end{eqnarray}
The terms occuring in the exponents are related to the quantum action.
Because the action is derived from a potential, energy is conserved. In imaginary time it reads
\begin{equation}
\label{eq:EnergCons}
E = - \tilde{T} + \tilde{V} = \mbox{const} ~ ,
\end{equation}
where $\tilde{T} = \frac{1}{2} \tilde{m} \dot{\tilde{x}}_{tr}^{2}$ denotes the kinetic term. Thus Eq.(\ref{eq:EnergCons}) can be resolved for the velocity,
\begin{equation}
\dot{\tilde{x}}_{tr} = \pm \sqrt{\frac{2}{\tilde{m}} ( \tilde{V}(\tilde{x}_{tr}) - E)} ~ .
\end{equation}
Then we can express the quantum action,
\begin{eqnarray}
\tilde{\Sigma}_{x_{in},0}^{x_{fi},T} &=& \int_{0}^{T} dt ~ \tilde{T} + \tilde{V} \nonumber \\
&=& \int_{0}^{T} E + 2 \tilde{T} = ET + \int_{0}^{T} dt ~ \tilde{m} \dot{\tilde{x}}_{tr}^{2}
\nonumber \\
&=& E T + \int_{x_{in}}^{x_{fi}} d\tilde{x} ~ \tilde{m} \dot{\tilde{x}}_{tr}
\nonumber \\
&=& E T + \int_{x_{in}}^{x_{fi}} dx ~ (\pm) \sqrt{2 \tilde{m} (\tilde{V}(x) - E)} ~ ,
\end{eqnarray}
where the sign is determined by the sign of the velocity $\dot{\tilde{x}}_{tr}$.

\begin{figure*}[ht]
\begin{center}
\includegraphics[scale=0.70,angle=0]{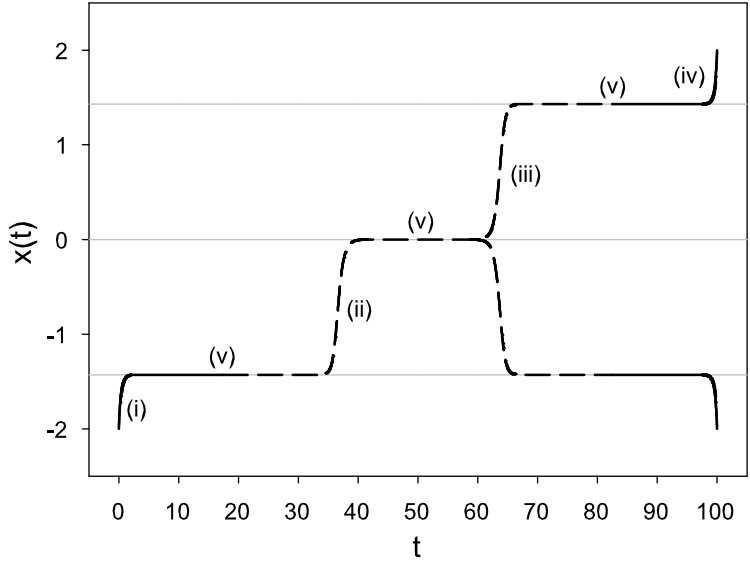}
\end{center}
\caption{Trajectory derived from quantum action $\tilde{\Sigma}$
for boundary conditions $x_{in}=-2$ and $x_{fi}=\pm 2$
($\lambda=\frac{1}{32}$ and $T=100$). Full line:
$\exp[-\tilde{\Sigma}]$. Dashed line: $\exp[+\tilde{\Sigma}]$.
Minima of quantum potential are indicated by thin gray lines. All quantities
are in dimensionless units.}
\label{fig:trajectory}
\end{figure*}

Now we consider the q.m. transition amplitude in imaginary time in the limit $T \to \infty$ (Feynman-Kac limit),
\begin{equation}
\label{eq:FeynmanKac}
G(y,T:x,0) \sim_{T \to \infty}
\psi_{gr}(y) \exp[-E_{gr}T/\hbar] \psi_{gr}(x) ~ .
\end{equation}
The coordinates $x$, $y$ may be located in any of the regimes I, II, III, IV.
For example let us consider $x,y \in I ~ (-\infty < x,y < -b)$.
Combining Eqs.(\ref{eq:FeynmanKac},\ref{eq:GroundState}) yields
\begin{eqnarray}
G(y,T:x,0) &=& Z^{2} ~ \exp[-E_{gr}T/\hbar]
\nonumber \\
&\times&
\exp \left[
- \int_{0}^{-b} ds \sqrt{\frac{2\tilde{m}}{\hbar^{2}} (\tilde{V}(s) - \tilde{V}_{0}) }
+ \int_{-b}^{y} ds \sqrt{\frac{2\tilde{m}}{\hbar^{2}} (\tilde{V}(s) - \tilde{V}_{0}) }
\right]
\nonumber \\
&\times&
\exp \left[
- \int_{x}^{-b} ds \sqrt{\frac{2\tilde{m}}{\hbar^{2}} (\tilde{V}(s) - \tilde{V}_{0}) }
+ \int_{-b}^{0} ds \sqrt{\frac{2\tilde{m}}{\hbar^{2}} (\tilde{V}(s) - \tilde{V}_{0}) }
\right] ~ .
\end{eqnarray}
There are five contributions in the exponent. Each of them can be
identified with the quantum action of some trajectory. For
example, consider the term
\begin{equation}
\exp \left[
- \int_{x}^{-b} ds \sqrt{ \frac{2 \tilde{m}}{\hbar^{2}} (\tilde{V}(s) - \tilde{V}_{0}) } \right] ~ .
\end{equation}
It corresponds to a trajectory starting from $x$ and approaching $-b$ some time later, i.e., entering the valley of the quantum potential at $-b$ (see Fig.[\ref{fig:trajectory}]).
Because the velocity $\dot{x} > 0$ along the trajectory, the corresponding sign in the quantum action is $+$. Thus the quantum action is
\begin{equation}
\tilde{\Sigma}_{x}^{-b} = + \int_{x}^{-b} ds
\sqrt{2 \tilde{m} (\tilde{V}(s) - \tilde{V}_0) } ~ .
\end{equation}
The term occuring in $G$ is
\begin{equation}
\exp \left[ - \int_{x}^{-b} ds \sqrt{ \frac{2 \tilde{m}}{\hbar^{2}} (\tilde{V}(s) - \tilde{V}_{0}) } \right]
= \exp[ - \frac{1}{\hbar} \tilde{\Sigma}_{x}^{-b} ] ~ ,
\end{equation}
which is valid, when we identify $E=\tilde{V}_{0}$. It gives a
minus sign in front of the action. Similarly, the other
contributions to $G$ can be expressed as some part of the
trajectory shown in Fig.[\ref{fig:trajectory}]. The term
\begin{equation}
\exp[-E_{gr}T/\hbar]
\end{equation}
corresponds to the trajectory, where the particle rests either in
the valley at $-b$ or in the valley at $0$ (having identified
$E_{gr}=\tilde{V}_{0}$).

To sum up, in the q.m. transition amplitude occur contributions
from five different trajectories. They are: trajectory from $x$ to
$-b$, then a straight line trajectory in the valley at $-b$, then
a trajectory going over from the valley at $-b$ to the valley at
$0$, then a straight line trajectory in the valley $0$, then a
trajectory going over from the valley $0$ to the valley $-b$, then
a straight-line trajectory in the valley $0$, and finally the
trajectory going from the valley $-b$ to the final boundary point
$y$. The transition amplitude then becomes
\begin{eqnarray}
G(y,T;x,0) = Z^{2}  \exp \left[ \frac{1}{\hbar} \left\{ - \tilde{\Sigma}_{x}^{-b}
+ \tilde{\Sigma}_{-b}^{0} + \tilde{\Sigma}_{0}^{-b} - \tilde{\Sigma}_{-b}^{y}
- \tilde{\Sigma}_{s.l.} \right\} \right] ~ .
\end{eqnarray}
Note that the straight line trajectory contribution can be taken
out of the action integral by computing the action in the
potential $\tilde{V}-\tilde{V}_0$ and adding $-E_{gr}T/\hbar$ to
the action afterwards. The full trajectory is shown in
Fig.[\ref{fig:trajectory}]. The full line corresponds the those
parts of the trajectory corresponding to the minus sign, while the
dashed line corresponds to parts with the plus sign. Let us take a
closer look at the trajectories shown in
Fig.[\ref{fig:trajectory}]. The trajectory going from $x=-2$ at
$t=0$ to $x=-2$ at $t=100$ has a contribution from an instanton
and an anti-instanton. Note that this trajectory does not minimize
the quantum action due to the contribution of the instanton
anti-instanton pair. This contribution is necessary to correctly
normalize the propagator. Other trajectories (corresponding to
other initial and final points) are built in the same way. If we
assign to each piece of trajectory a corresponding sign in front
of the quantum action, we can write
\begin{eqnarray}
G(y,T;x,0) = Z^{2}  \exp\left[ \frac{1}{\hbar} \sum_{traj(\nu)} ~ \mbox{sgn}_{traj(\nu)} ~ \tilde{\Sigma}_{traj(\nu)} \right] ~ .
\end{eqnarray}
For other regimes, where the quantum potential has only one well,
the correct trajectory is much simpler to find and all of it is
accounted negatively in the action.

\section{Numerical simulations}
\label{sec:NumSim} In this section we want to present numerical
simulations of the quantum action and see how well it fits the
transition amplitudes. The computation of the quantum action
functional requires to compute a trajectory. Such trajectory is a
solution of the Euler-Lagrange equation of motion (in imaginary
time), and satisfies boundary conditions at initial and final
points. The numerical solution of such differential equations has
been found to be most convenient and give stable results by using
a relaxation algorithm. To give an example how we have proceeded,
consider Fig.[\ref{fig:trajectory}], in particular, the trajectory
going from $x=-2$ to $x=-2$. Let us recall that we work in the
regime where the transition time $T$ is large. In this regime we
observe that the trajectory has three pieces, where its motion
follows the bottom of a potential valley (first from $t=5$ to
$t=35$,  second from $t=45$ to $t=60$, finally from $t=70$ to
$t=95$). Each of those pieces of trajectory in the valley can be
cut somewhere in the middle. As a result the whole trajectory is
decomposed in four parts; first from $t=0$ to $t=20$, second from
$t=20$ to $t=50$, third from $t=50$ to $t=80$, and fourth from
$t=80$ to $t=100$. Each of those four pieces can be computed
separately in a numerical way. Finally, the assignment of the sign
of the quantum action is taken from the above theoretical
analysis.

\begin{figure*}[ht]
\begin{center}
\includegraphics[scale=0.70,angle=0]{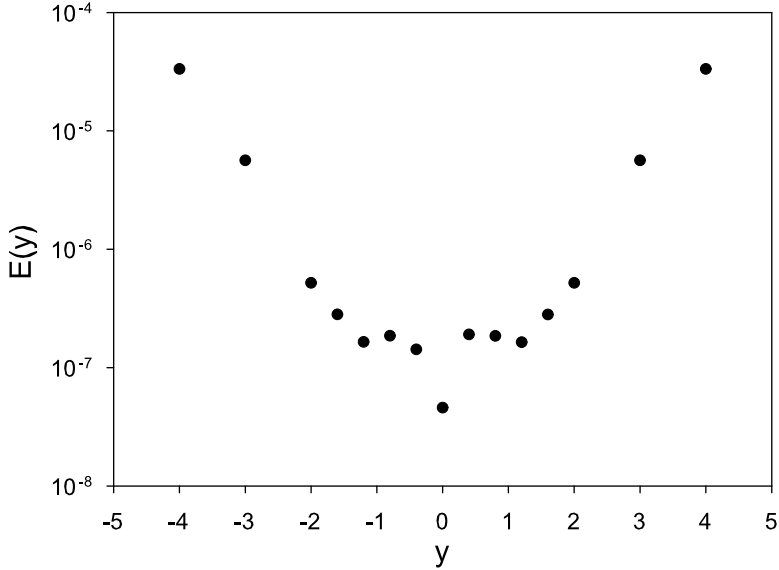}
\end{center}
\caption{Relative error $E(y)=\left\vert\frac{
G(y,T:0,0)_{QA}-G(y,T:0,0)_{Schrod}}
{G(y,T:0,0)_{Schrod}}\right\vert$ of transition amplitude for
various final positions $y$ for $T=100$ and
$\lambda=\frac{1}{32}$. All quantities are in dimensionless units.}
\label{fig:ErreurAmplitude}
\end{figure*}

The system provides two quantities that can be used to find a characteristic time scale, namely the natural frequency of oscillation in each well ($T_{scale}^{(1)} = \omega^{-1} = 1$) and the ground state energy of the system ($T_{scale}^{(2)} = \hbar/E_{gr}$). Since $E_{gr}$ is in the order of 1 in the range of parameters considered in this paper (and since we work in units where $\hbar=1$), we simply take $T_{scale} = 1$. We chose to study transitions at times $T=10$ and $T=100$, which are both large compared to the dynamical time scale. For large transition time $T \gg T_{scale}$ the transition amplitude is dominated by the ground state and a few excited states. The wave functions and energies of those states have been computed by numerically solving the Schr\"odinger equation.
The quantum action, Eq.(\ref{eq:DefQAction}), is determined by the parameter of the quantum mass $\tilde{m}$ and the quantum potential $\tilde{V}(x)$. We computed the function $\tilde{m}(\tilde{V}(x)- \tilde{V}_{0})$ from Eq.(\ref{eq:DefQPot}). In order to determine the quantum mass we need another equation.
\begin{figure*}[ht]
\begin{center}
\includegraphics[scale=0.70,angle=0]{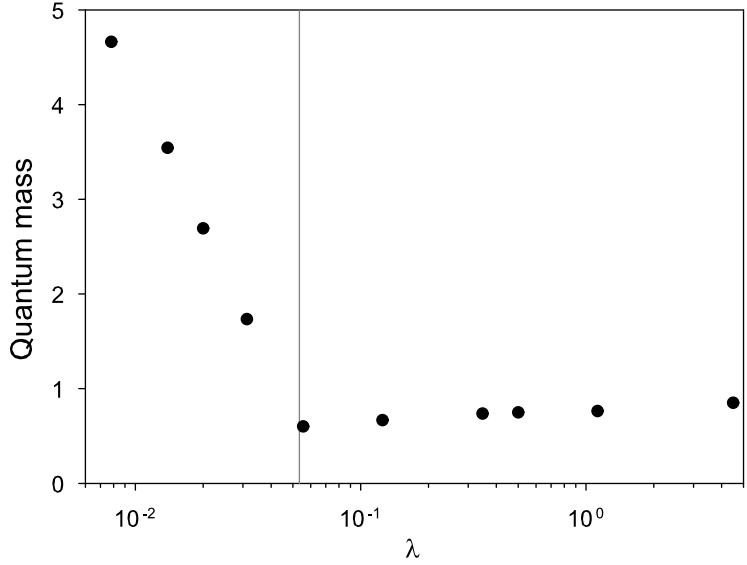}
\end{center}
\caption{Quantum mass vs. parameter $\lambda$. $T=10$. All quantities
are in dimensionless units.}
\label{fig:QMass}
\end{figure*}
In the Feynman-Kac limit, one cannot directly find the quantum mass independently from the quantum potential. The physical reason is that in the Feynman-Kac limit physics is dominated by the ground state. According to Eq.(\ref{eq:WaveFct}), in the ground state wave function occurs the product of quantum mass and quantum potential. The mathematical reason is that there is a symmetry transformation \cite{Q4}, which keeps the transition amplitude and the quantum action invariant.
The transformation
\begin{eqnarray}
&& m \to m'= m/\alpha
\nonumber \\
&& V(x) \to V'(x) = \alpha V(x)
\nonumber \\
&& T \to T' = T/\alpha
\end{eqnarray}
leaves the transition amplitudes $G(y,T;x,0)$ invariant.
The transformation
\begin{eqnarray}
&& \tilde{m} \to \tilde{m}'= \tilde{m}/\alpha
\nonumber \\
&& \tilde{V}(x) \to \tilde{V}'(x) = \alpha \tilde{V}(x)
\nonumber \\
&& T \to T' = T/\alpha ~ ,\label{eq:sym}
\end{eqnarray}
leaves the quantum action $\tilde{\Sigma}_{x,0}^{y,T}$ invariant.
This invariance implies that the choice of the quantum mass
$\tilde{m}$ is arbitrary when T is very large: this symmetry
strictly holds in the limit $T \to \infty$. Numerically, we found
that this freedom in the choice of $\tilde{m}$ was valid to good
precision for $T \ge 100$, that is, the value of the quantum mass
had no significant influence on the transition amplitudes for
$T=100$. However, this was found not to be the case for $T=10$.
Therefore, we need additional information about the system to find
the correct quantum mass for $T=10$. We have proceeded in the
following way. For a given value of $\lambda$ we made an initial
guess of the quantum mass $\tilde{m}$. Then we computed the
quantum action for a number of initial points $x_{in}$ and final
points $x_{fi}$, taken from a set $\{x_1, \dots, x_{N}\}$. Thus we
generated a matrix of quantum action elements $\Sigma_{i}^{j}$.
Via the quantum action functional, $G_{ij} = Z
\exp[\Sigma_{i}^{j}]$, we computed a matrix of transition matrix
elements. Diagonalisation of $G_{ij}$ yields eigenvalues
$E_{n}T/\hbar$. Then we made a variational search in the parameter
$\tilde{m}$, until the energy of the first excited state $E_{1}$,
obtained from diagonalisation of $G_{ij}$ agreed with the exact
value (obtained from the solution of the Schr\"odinger equation).
On the other hand, previous numerical experiments have shown that
the quantum mass (and also the parameters of the quantum
potential) asymptotically converge when $T \to \infty$. Therefore,
we simply chose the same values of the quantum mass in the $T=100$
case. The obtained results for the quantum mass are shown in
Fig.[\ref{fig:QMass}] as function of the parameter $\lambda$.
One observes for large $\lambda$ ($\lambda > 6 \cdot 10^{-2}$) a smooth behavior of $\tilde{m}$. At $\lambda \approx 5.3 \cdot 10^{-2}$ there is a cusp.
In our opinion the behavior for $\lambda < 5 \cdot 10^{-2}$ is unphysical, caused by limited numerical precision and the fact that in this regime the contribution to the propagator from the first excited state becomes non-negligeable.
However, this uncertainty does not translate into a large error on the transition amplitudes because of their vanishing dependence on the quantum mass in the limit $T \to \infty$.
The relative error of the transition amplitude
as a function of final position $y$, while keeping the initial
position fixed ($x=0$), is shown in
Fig.[\ref{fig:ErreurAmplitude}] for $T=100$ and
$\lambda=\frac{1}{32}$. The error is larger for large $y$ because
the numerical error on the quantum potential is larger for large
$\vert x\vert$.
\begin{figure*}[ht]
\begin{center}
\includegraphics[scale=0.70,angle=0]{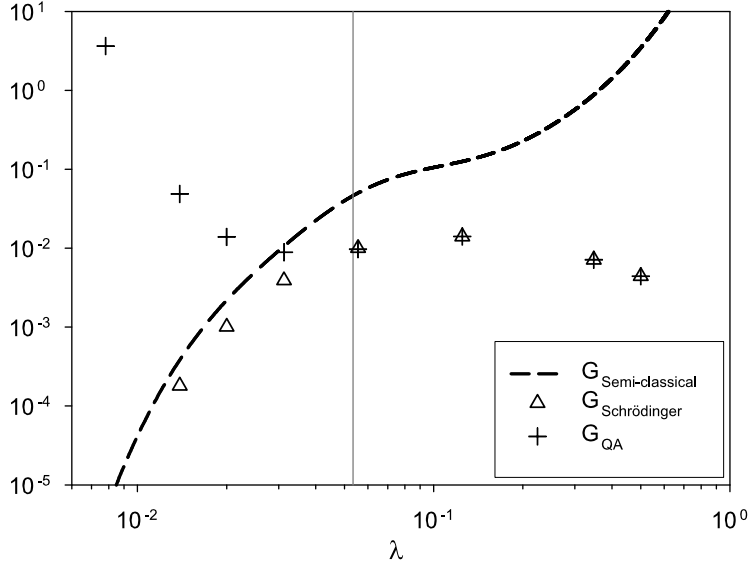}
\end{center}
\caption{Tunneling amplitude $G(a,T;-a,0)$ vs.
potential parameter $\lambda$. $T=10$. Comparison
of exact result (from Schr\"odinger eq.) with result from quantum
action functional and from instanton method (semi-classical). 
All quantities are in dimensionless units.}
\label{fig:Tunn10}
\end{figure*}

\section{Tunneling: Comparison with instanton picture }
\label{sec:CompInst}
\begin{figure*}[ht]
\begin{center}
\includegraphics[scale=0.70,angle=0]{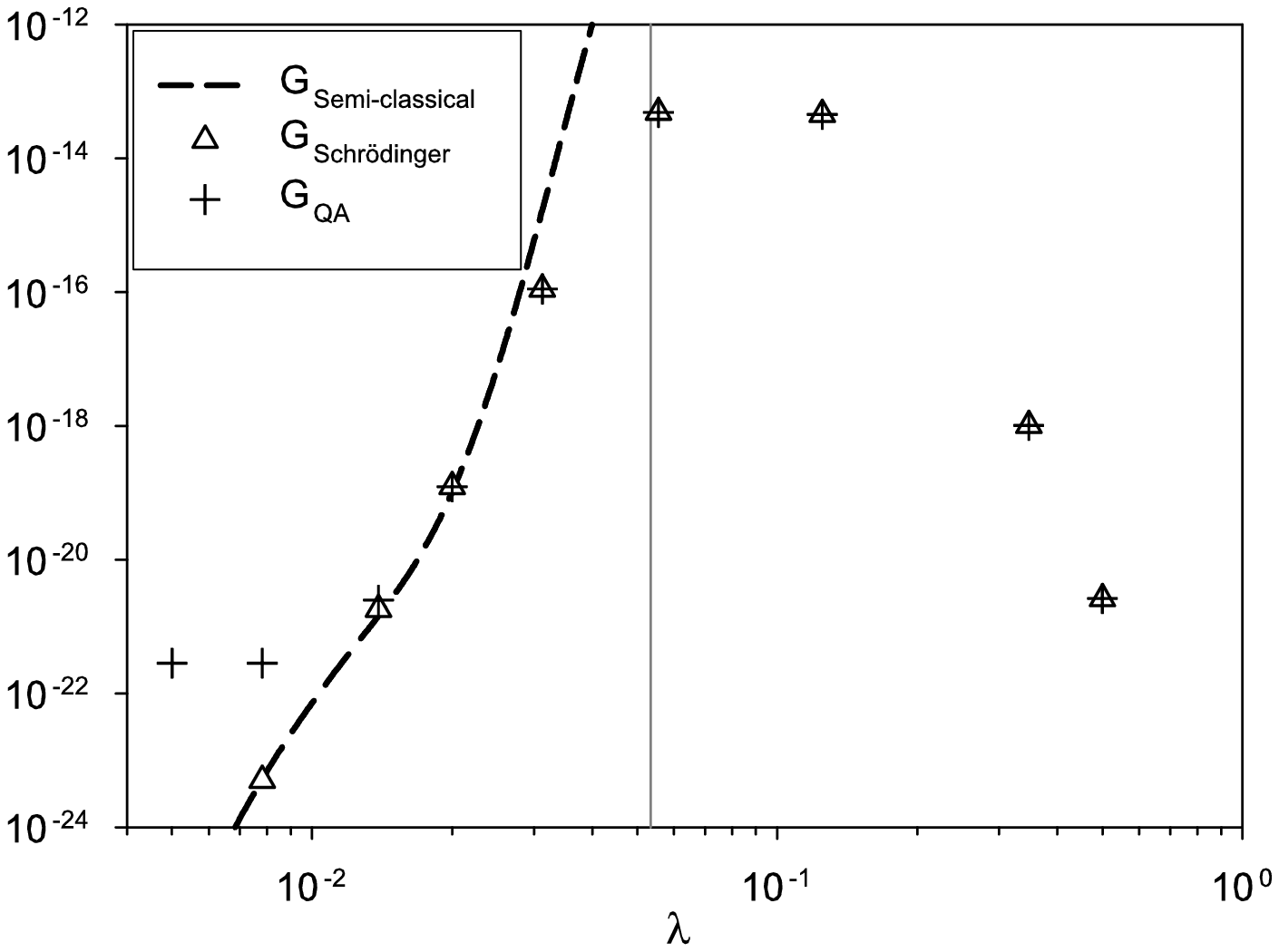}
\end{center}
\caption{Same as Fig.[\ref{fig:Tunn10}] but $T=100$.
All quantities are in dimensionless units.}
\label{fig:Tunn100}
\end{figure*}

\begin{figure*}[ht]
\begin{center}
\includegraphics[scale=0.70,angle=0]{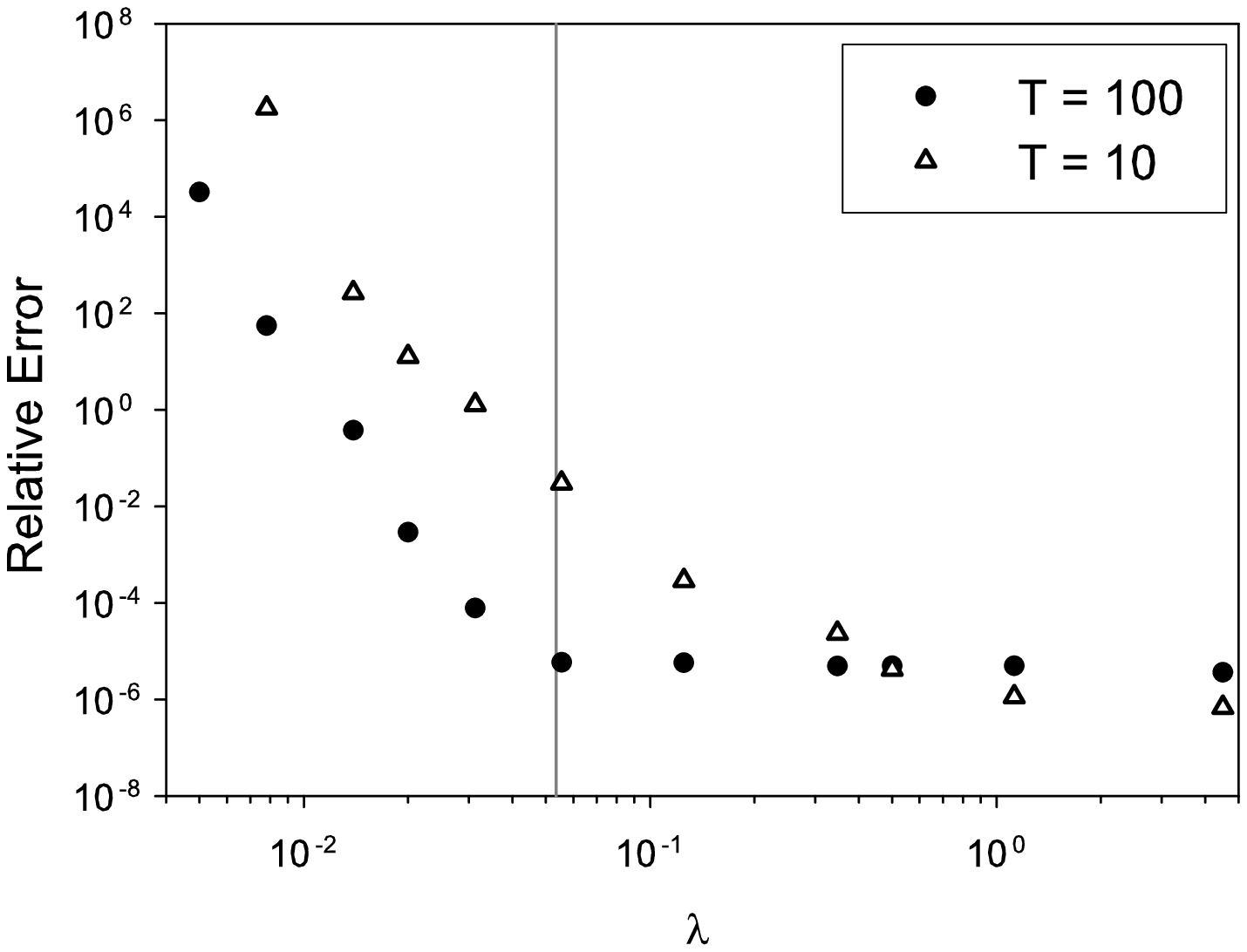}
\end{center}
\caption{Relative error of the tunneling amplitude from the
quantum action functional with respect to exact result, for $T=10$
and $T=100$. All quantities are in dimensionless units.}
\label{fig:Erreur}
\end{figure*}

We have determined the tunneling amplitude
\begin{equation}
G_{Tun} \equiv G(a,T:-a,0) =
\langle a\vert \exp [-HT/\hbar] \vert -a \rangle ~ ,
\end{equation}
where $\pm a$ are the minima of the classical potential, for
various values of the potential parameter $\lambda$ and transition
time $T$. As reference value, the tunneling amplitude has been
computed by solving the Schr\"odinger equation. Next, the tunneling
transition amplitude has been computed from the quantum action
functional, that is by computing the quantum potential and the
quantum mass for each of the different classical potentials and
transition times and by using the tools developed in
Sect.\ref{sec:ModelAndAction}. Moreover, the transition amplitudes
have been obtained by using the semi-classical multi-instanton
expression at two-loop order, given by Ref.\cite{Schaefer},
\begin{equation}
G_{Tun} = \sqrt{\frac{\omega}{\pi}}\left( 1 +
\frac{3}{8S_0}\right) \exp\left( -\frac{\omega T}{2} \left[
1-\frac{1}{3S_0}\right] \right) \sinh{\left(
\sqrt{\frac{6S_0}{\pi}}\exp\left[-S_0-\frac{71}{72}\frac{1}{S_0}\right]\omega
T\right) } ~ ,
\end{equation}
where $\hbar = 1$ and $S_0 \equiv 1/(12\lambda)$ is the action of
the instanton between classical minima. Results are shown in
Figs.[\ref{fig:Tunn10},\ref{fig:Tunn100}] for $T=10$ and $T=100$,
respectively. The relative difference between the exact tunneling
amplitude (from Schr\"odinger equation) and the quantum action
amplitude for $T=10$ and $T=100$ is shown in
Fig.[\ref{fig:Erreur}].
The instanton method is valid in the semiclassical limit where the
action $S_0$ of the classical instanton is large compared to unity
(corresponding to small $\lambda$). This is clearly the case in
Figs[\ref{fig:Tunn10},\ref{fig:Tunn100}] where the semi-classical
tunneling amplitude is quite close to the exact amplitude for
small $\lambda$. In contrast, the tunneling amplitude computed
from the quantum action is better at large $\lambda$. It diverges
substantially from the exact result for small $\lambda$. This
divergence is \emph{not} due to a bad choice of quantum mass. It
can rather be explained by the fact that the quantum action is
constructed from the exact propagator \emph{in the Feynman-Kac
limit}, where only the ground state contributes to the physics of
the system. However, when $\lambda$ decreases, the wells get
deeper and the energy of the first excited approaches the ground
state energy. Then the first excited state contribution cannot be
neglected anymore (the ground state becomes degenerate). We can
verify that the quantum action functional indeed reproduces
correctly the ground state contribution to the propagator
$\vert\psi_{gr}\rangle e^{-E_{gr}T/\hbar}\langle\psi_{gr}\vert$
for all $\lambda$, up to numerical limits. The relative error of
tunneling amplitudes from the quantum action is shown in
Fig.[\ref{fig:Erreur}]. The fact that the error is generally
higher for $T=10$ confirms this observation (lower transition time
means that the first excited state is more important in the
transition amplitude for the same value of $\lambda$). It turns
out that the quantum action can be used to compute transition
amplitudes for a wide range of potential parameters.
\begin{figure*}[ht]
\begin{center}
\includegraphics[scale=0.70,angle=0]{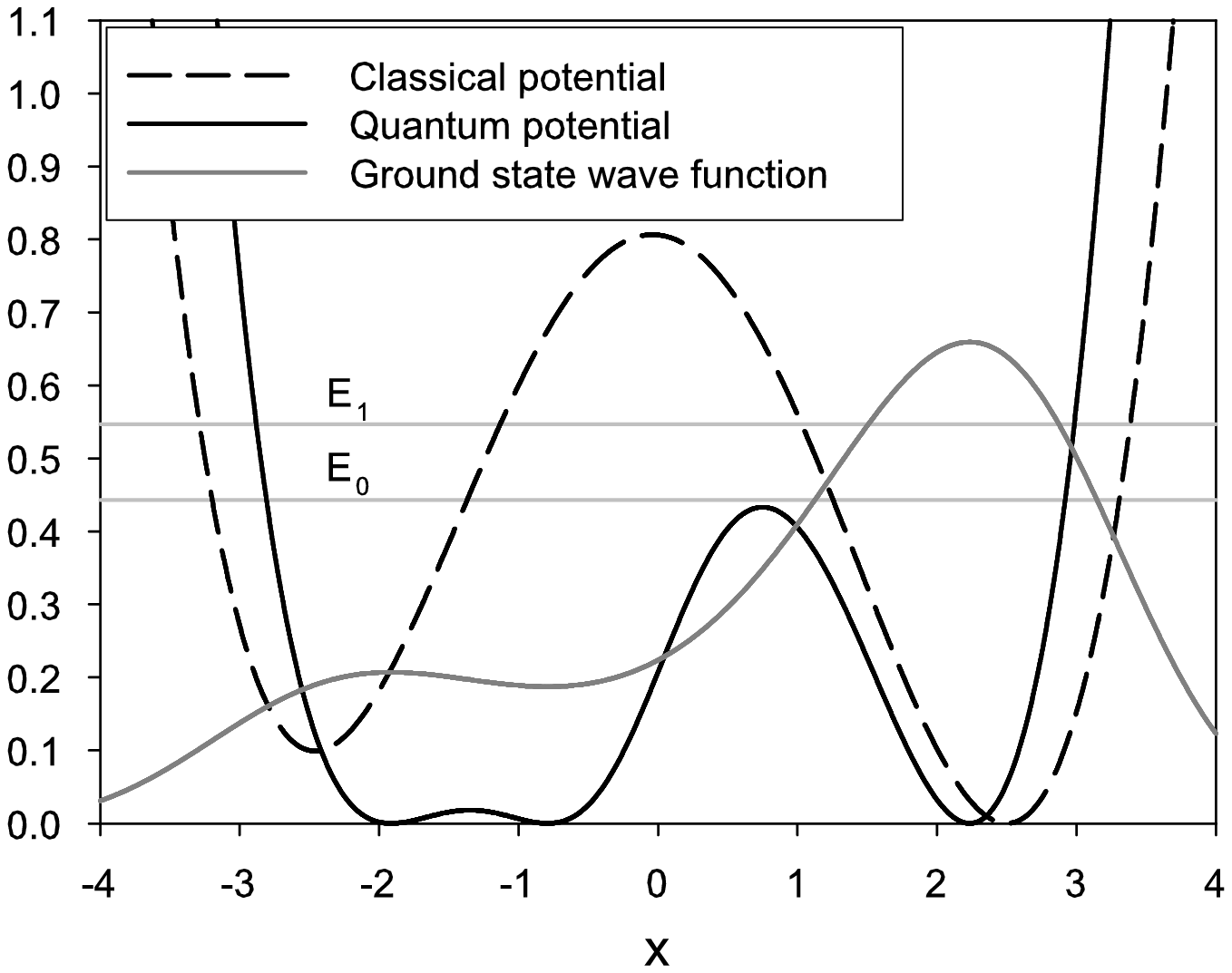}
\end{center}
\caption{Asymmetric double-well potential, Eq.(\ref{eq:AsymPot}),
the ground state wave function and corresponding quantum potential
$\tilde{V}-\tilde{V}_0$ (up to a multiplicative factor). All quantities are in dimensionless units.}
\label{fig:AsymPot}
\end{figure*}

\section{Tunneling in asymmetric potentials}
\label{sec:tilted} Another advantage of the quantum action
functional is that it can be used to study tunneling in an
asymmetric double-well potential, with wells of different depth.
While such a system is much harder to study in the standard
instanton picture, it does not require substantially more work in
the quantum action context. In fact, the quantum potential will
still display one, two or three minima, \emph{having all the same
depth}. Instantons trajectories similar to those in the standard
double-well case then appear likewise. For example, consider the
potential
\begin{equation}
\label{eq:AsymPot}
V(x) = \frac{1}{50}\left[x^2-\left(\frac{5}{2}\right)^2\right]^2+\frac{1}{250}\left[x-\frac{5}{2}\right]^2,
\end{equation}
which is shown in Fig.[\ref{fig:AsymPot}]. This potential has two
wells, but the left well is higher than the right one, and
therefore a classical instanton does not exist. The quantum
potential (up to a multiplicative factor) recovered from the
ground state wave function is also shown in
Fig.[\ref{fig:AsymPot}]. We find that the quantum mass in this
case (for $T=100$) is about $\tilde{m}=0.11$ and that the
tunneling transition amplitude $G(a,T;-a,0)$ reproduces the exact
value with a relative error of $2.87\cdot 10^{-4}$.

\section{Discussion}
\label{sec:Discuss} Using the quantum action functional to
describe tunneling in a double well potential gives good results
in the deep quantum regime (opposite to the semi-classical
regime). Quantum effects appear in the quantum action via tuned
parameters. In the case of tunneling in a classical double-well
potential this leads to a quantum potential with a different well
structure, having possibly one, two or three wells. Like in the
standard instanton picture, quantum instantons occur also in the
tunneling amplitude obtained by the quantum action functional.
Instantons play an important role in many domains of physics so
one may ask how important are the instantons in this case. First,
we saw that they play a normalization role in the trajectories,
that is, they ensure that the propagator is correctly normalized.
Second, as much as standard instantons represent the only finite
action solution (in the limit $T\to\infty$) contributing to the
propagator between the classical minima, we can interpret the new
instantons as the only finite action solutions (in the
$T\to\infty$ limit) contributing to the propagator \emph{between
the minima of the quantum potential}. Because these minima of the
quantum potential correspond in position to the extrema of the
ground state wave function, the double quantum instanton going
from $-b$ to $0$ to $b$ represents the trajectory reproducing the
largest transition amplitude. In particular, the tunneling
amplitude between the minima of the quantum potential $\langle
-b\vert \exp[-HT/\hbar]\vert b \rangle$ is larger than the
tunneling amplitude between the minima of the classical potential
$\langle -a\vert \exp[-HT/\hbar]\vert a \rangle$. Third, one might
wonder why in the quantum action functional only two instantons,
respectively anti-instantons, contribute, while in the standard
approach infinitely many instantons and anti-instantons are
needed. The answer is simply that the quantum action was
explicitely constructed to reproduce \emph{exactly}, in the
Feynman-Kac limit, the transition amplitude, using a single
trajectory. This is sufficient to discard the use of multiple
instanton trajectories. Fourth, let us compare the structure of
the classical instanton with the quantum instanton. The instanton
solution corresponding to the classical Hamiltonian of
Eqs.(\ref{eq:Hamilton},\ref{eq:Potential}) is given by
\begin{equation}
\label{eq:InstSol} x_{inst}(t) = \frac{1}{\sqrt{8 \lambda}} ~
\mbox{tgh} \left[ \frac{1}{\sqrt{4m}} (t-t_c) \right] = \sqrt{8 B}
~ \mbox{tgh} \left[ \frac{1}{\sqrt{4m}} (t-t_c) \right] ~ ,
\end{equation}
where $B$ denotes the barrier height of the potential and $t_c$ is
the center of the instanton. The steepness of the instanton at its
center is given by
\begin{equation}
\label{eq:InstSteep} \frac{d}{dt} x_{inst}(t=t_c) = \sqrt{
\frac{2B}{m} } ~ .
\end{equation}
\begin{figure*}[ht]
\begin{center}
\includegraphics[scale=0.70,angle=0]{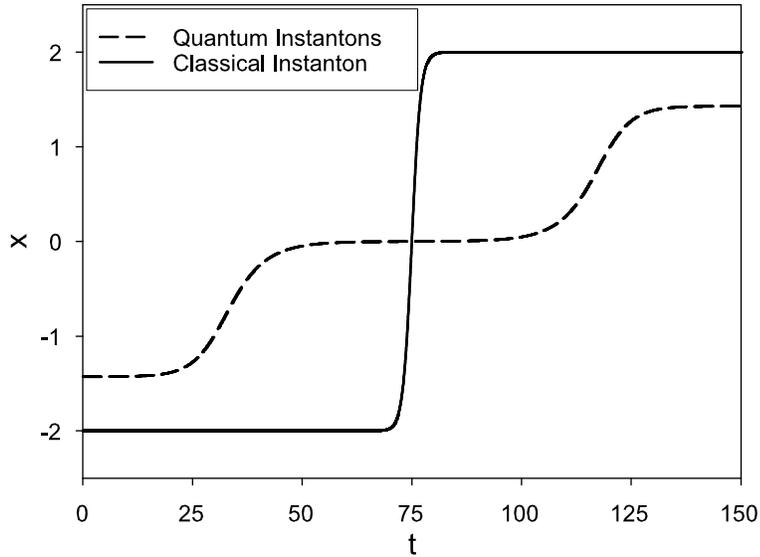}
\end{center}
\caption{Comparison between the classical and quantum instantons,
for $\lambda = \frac{1}{32}$. All quantities are in dimensionless units.}
\label{fig:CompInst}
\end{figure*}
Fig.[\ref{fig:PotL1-32}] shows that that the barrier height of the
quantum potential is much smaller than that of the classical
potential. The quantum mass is of the same order of magnitude as
the classical mass. From Eq.({\ref{eq:InstSteep}) we expect that
the steepness of the quantum instanton is smaller than that of the
classical instanton. This is confirmed by a numerical calculation,
comparing the classical with the quantum instanton, shown in
Fig.[\ref{fig:CompInst}]. Also because the location of the
potential minima are closer for the quantum potential than for the
classical potential, the quantum instanton has a smaller action
than the classical one. The bottom line is that the quantum
instanton is "softer" than the classical instanton. It is
interesting to note that a similar observation has been made
previously in the context of comparing classical chaos with
quantum chaos, again using the quantum action functional
\cite{Caron04}. It has been found that the quantum action yields a
less chaotic phase space than the classical action. The underlying
reason for such behavior is unknown to us. We believe that a
promising strategy may be to analyze the path integral and
its relation to the quantum action functional.

\section{Summary}
\label{sec:Summary} This work is about tunneling described in
terms of the quantum action functional. This point of view is
complementary to the standard instanton picture: While the latter
holds in the semi-classical regime, the former holds in the deep
quantum regime. The observation that the quantum potential has new
minima (in number and location) beyond those of the classical
potential may be of interest for cosmology and inflationary
models. There are experiments on tunneling in condensed matter,
e.g. Josephson junctions, SQUIDS \cite{Friedman00,Averin00} or in
atomic physics in dynamical tunneling of atoms in a time-dependent
exterior field \cite{Hensinger,Raizen}. It would be interesting to
explore if the quantum action functional can be applied to
describe such physics. This requires further development, in
particular, to explore the quantum action functional in real time
and for explicitely time-dependent systems.
Likewise one may ask if tunneling out of meta-stable states can be described in our approach. We have shown that an asymmetric double-well potential which may give rise to meta-stable states (which occur, e.g., in nuclear fission and emission of $\alpha$-particles) can be treated in this framework, provided that the potential is bounded from below (which excludes a potential like $V \sim x^{3}$). However, the description of tunneling from quasi-stationary states to the ground state or other excited states would require to apply the quantum action in real time. This will be a subject of further studies.

\vspace{0.5cm}
\noindent {\bf Acknowledgements} \\
F.P., H.K. and K.M. have been supported by NSERC Canada.
H.K. is very grateful to E. Shuryak and T. Sch\"afer for discussions.

\end{document}